\documentclass[12pt]{article}
\usepackage{epsfig}
\usepackage{psfrag}
\usepackage{latexsym}
\usepackage{indentfirst}
\usepackage{fancyhdr}
\usepackage{amssymb}
\usepackage{amsfonts}
\usepackage{cite}
\usepackage{bbold}
\usepackage[footnotesize]{caption2}
\usepackage{graphicx}
\usepackage[center,footnotesize,hang]{subfigure}
\usepackage{url}

\newcommand{\be}{\beta}

\newcommand{\beq}{\begin{equation}}
\newcommand{\eeq}{\end{equation}}
\newcommand{\bac}{\beq\begin{array}}
\newcommand{\eac}{\end{array}\eeq}
\newcommand{\ba}{\begin{array}}
\newcommand{\ea}{\end{array}}
\newcommand{\bea}{\begin{eqnarray}}
\newcommand{\eea}{\end{eqnarray}}
\newcommand{\beaa}{\begin{eqnarray*}}
\newcommand{\eeaa}{\end{eqnarray*}}

\newcommand{\nn}{\nonumber}

\def\beq{\begin{equation}}
\def\eeq{\end{equation}}
\def\bea{\begin{eqnarray}}
\def\eea{\end{eqnarray}}
\def\bet{\begin{tabular}}
\def\eet{\end{tabular}}
\def\bes{\begin{subequations}\bea}
\def\ees{\eea\end{subequations}}

\def\be{\begin{equation}}
\def\ee{\end{equation}}
\def\bc{\begin{center}}
\def\ec{\end{center}}
\def\bea{\begin{eqnarray}}
\def\eea{\end{eqnarray}}
\def\dd{\displaystyle}
\def\nn{\nonumber}

\catcode`@=11
\def\marginnote#1{}
\newcount\hour
\newcount\minute
\newtoks\amorpm
\hour=\time\divide\hour by60
\minute=\time{\multiply\hour by60 \global\advance\minute by-\hour}
\edef\standardtime{{\ifnum\hour<12 \global\amorpm={am}%
        \else\global\amorpm={pm}\advance\hour by-12 \fi
        \ifnum\hour=0 \hour=12 \fi
        \number\hour:\ifnum\minute<10 0\fi\number\minute\the\amorpm}}
\edef\militarytime{\number\hour:\ifnum\minute<10 0\fi\number\minute}
\def\draftlabel#1{{\@bsphack\if@filesw {\let\thepage\relax
   \xdef\@gtempa{\write\@auxout{\string
      \newlabel{#1}{{\@currentlabel}{\thepage}}}}}\@gtempa
   \if@nobreak \ifvmode\nobreak\fi\fi\fi\@esphack}
        \gdef\@eqnlabel{#1}}
\def\@eqnlabel{}
\def\@vacuum{}
\def\draftmarginnote#1{\marginpar{\raggedright\scriptsize\tt#1}}
\def\draft{\oddsidemargin 0.0truein
        \def\@oddfoot{\sl preliminary draft \hfil
        \rm\thepage\hfil\sl\today\quad\militarytime}
        \let\@evenfoot\@oddfoot \overfullrule 3pt
        \let\label=\draftlabel
        \let\marginnote=\draftmarginnote
   \def\@eqnnum{(\theequation)\rlap{\kern\marginparsep\tt\@eqnlabel}%
\global\let\@eqnlabel\@vacuum}  }
\catcode`@=12
\begin{document}
\begin{titlepage}
\vspace*{-1cm}
\phantom{hep-ph/***}

\hfill{RM3-TH/09-8}
\hfill{CERN-PH-TH/2009-046}

\vskip 2.5cm
\begin{center}
{\Large\bf A Simplest A4 Model for \\ Tri-Bimaximal Neutrino Mixing }
\end{center}
\vskip 0.2  cm
\vskip 0.5  cm
\begin{center}
{\large Guido Altarelli}~\footnote{e-mail address: guido.altarelli@cern.ch}
\\
\vskip .1cm
Dipartimento di Fisica `E.~Amaldi', Universit\`a di Roma Tre
\\
INFN, Sezione di Roma Tre, I-00146 Rome, Italy
\\
\vskip .1cm
and
\\
CERN, Department of Physics, Theory Division
\\
CH-1211 Geneva 23, Switzerland
\\

\vskip .2cm
{\large Davide Meloni}~\footnote{e-mail address: meloni@fis.uniroma3.it} \\
\vskip .1cm
Dipartimento di Fisica `E.~Amaldi', Universit\`a di Roma Tre
\\
INFN, Sezione di Roma Tre, I-00146 Rome, Italy
\\
\end{center}
\vskip 0.7cm
\begin{abstract}
\noindent
We present a see-saw $A_4$ model for Tri-Bimaximal mixing which is based on a very economical flavour symmetry and field content and still possesses all the good features of $A_4$ models. In particular the charged lepton mass hierarchies are determined by the $A_4\times Z_4$  flavour symmetry itself without invoking a Froggatt-Nielsen  $U(1)$ symmetry. Tri-Bimaximal mixing is exact in leading order while all the mixing angles receive corrections of the same order in next-to-the-leading approximation.  As a consequence the predicted value of $\theta_{13}$ is within the sensitivity of the experiments which will take data in the near future. The light neutrino spectrum, typical of $A_4$ see-saw models, with its phenomenological implications, also including leptoproduction, is studied in detail.
\end{abstract}
\end{titlepage}
\setcounter{footnote}{0}
\vskip2truecm
%
\section{Introduction}
It is an experimental fact \cite{data1,data,FogliIndication,MaltoniIndication} that within measurement errors
the observed neutrino mixing matrix \cite{review} is compatible with
the so called Tri-Bimaximal (TB) form \cite{hps}. The best measured neutrino mixing angle $\theta_{12}$ is just about 1$\sigma$ below the TB value $\tan^2{\theta_{12}}=1/2$, while the other two angles are well inside the 1$\sigma$ interval \cite{data}. In a series of papers \cite{TBA4,AFextra,AFmodular,AFL, AFH} it has been pointed out that a broken flavour symmetry based on the discrete
group $A_4$ appears to be particularly suitable to reproduce this specific mixing pattern in leading order (LO). Other
solutions based on alternative discrete or  continuous flavour groups have also been considered \cite{continuous,others,bmm}, but the $A_4$ models have a very economical and attractive structure, e.g. in terms of group representations and of field content. In most of the models $A_4$ is accompanied by additional flavour symmetries, either discrete like $Z_N$ or continuous like U(1), which are necessary to eliminate unwanted couplings, to ensure the needed vacuum alignment and to reproduce the observed charged lepton mass hierarchies.  Given the set of flavour symmetries and having specified the field content, the non leading corrections to the TB mixing arising from loop effects and higher dimensional operators can be evaluated in a well defined expansion. In the absence of specific dynamical tricks, in a generic model, all the three mixing angles receive corrections of the same order of magnitude. Since the experimentally allowed departures of $\theta_{12}$ from the TB value $\sin^2{\theta_{12}}=1/3$ are small, at most of $\mathcal{O}(\lambda_C^2)$, with $\lambda_C$ the Cabibbo angle, it follows that both $\theta_{13}$ and the deviation of $\theta_{23}$ from the maximal value are expected in these models to also be at most of $\mathcal{O}(\lambda_C^2)$ (note that $\lambda_C$ is a convenient hierarchy parameter not only for quarks but also in the charged lepton sector with $m_\mu/m_\tau \sim0.06 \sim \mathcal{O}(\lambda_C^2)$ and $m_e/m_\mu \sim 0.005\sim\mathcal{O}(\lambda_C^{3-4})$). A value of $\theta_{13} \sim \mathcal{O}(\lambda_C^2)$ is within the sensitivity of the experiments which are now in preparation and will take data in the near future. In this paper we present an $A_4$ model for TB mixing which is based on a most economical flavour symmetry and field content and still possesses all the features described above. In particular TB mixing is exact in LO while all mixing angles receive corrections at higher orders. The charged lepton mass hierarchies are determined by the $A_4\times Z_4$ flavour symmetry itself without invoking a Froggatt-Nielsen  $U(1)$ symmetry, as a consequence of a particular alignment as proposed in refs. \cite{lin1,lin2}. Our model, which is of the see-saw type, differs from those in refs.\cite{lin1,lin2} because the flavour symmetry is smaller and the pattern of corrections to TB mixing is more general and flexible. It is interesting that $A_4$ models with the see-saw mechanism typically lead to a light neutrino spectrum which satisfies the sum rule (among complex masses):
\be
\frac{1}{m_3}=\frac{1}{m_1}-\frac{2}{m_2}~.\\
\label{sumr}
\ee
We discuss the features of the spectrum in detail, as these properties can be considered as a signal for confirming the underlying $A_4$ symmetry.

The article is organized as follows.  In sect. 2 the structure of the model is described. In sect. 3 we summarize the contributions from next to the leading order corrections. In sects. 4-6 the phenomenological consequences are discussed.  Finally, sect. 7 is devoted to our conclusion.

%
\section{The structure of the model}

We introduce here the structure of our model,  which leads to  TB mixing in first approximation. The model is formulated in terms of the $A_4$  realization in the T diagonal basis introduced in ref.\cite{AFmodular}. We recall that $A_4$, the group of even permutations of 4 objects, can be generated by the two elements
$S$ and $T$ obeying the relations (a "presentation" of the group):
\be
S^2=(ST)^3=T^3=1~.
\label{$A_4$}
\ee
The 12 elements of $A_4$  are obtained as:
$1$, $S$, $T$, $ST$, $TS$, $T^2$, $ST^2$, $STS$, $TST$, $T^2S$, $TST^2$, $T^2ST$.
The inequivalent irreducible representations of $A_4$ are 1, $1'$, $1''$ and 3. It is immediate to see that one-dimensional unitary representations are
given by:
\be
\begin{array}{lll}
1&S=1&T=1\\
1'&S=1&T=e^{\dd i 4 \pi/3}\equiv\omega^2\\
1''&S=1&T=e^{\dd i 2\pi/3}\equiv\omega \,.\\
\label{s$A_4$}
\end{array}
\ee
The three-dimensional unitary representation, in a basis
where the element $T$ is diagonal, is given by:
\be
T=\left(
\begin{array}{ccc}
1&0&0\\
0&\omega^2&0\\
0&0&\omega
\end{array}
\right),~~~~~~~~~~~~~~~~
S=\frac{1}{3}
\left(
\begin{array}{ccc}
-1&2&2\cr
2&-1&2\cr
2&2&-1
\end{array}
\right)~.
\label{ST}
\ee
It is useful to remind the product rules of two triplets, ($\psi_1,\psi_2,\psi_3$) and ($\varphi_1,\varphi_2,\varphi_3$) of $A_4$, according to the multiplication rule 3x3=$1+1'+1''+3_A+3_S$:
\bea \label{tensorproda4}
&\psi_1\varphi_1+\psi_2\varphi_3+\psi_3\varphi_2 \sim 1 ~,\nn \\
&\psi_3\varphi_3+\psi_1\varphi_2+\psi_2\varphi_1 \sim 1' ~,\nn \\
&\psi_2\varphi_2+\psi_3\varphi_1+\psi_1\varphi_3 \sim 1'' ~,\nn
\eea
  \be
   \left( 
 \ba{c}
2\psi_1\varphi_1-\psi_2\varphi_3-\psi_3\varphi_2 \\
2\psi_3\varphi_3-\psi_1\varphi_2-\psi_2\varphi_1 \\
2\psi_2\varphi_2-\psi_1\varphi_3-\psi_3\varphi_1 \\
  \ea
  \right) \sim 3_S~, \qquad
  \left( 
 \ba{c}
\psi_2\varphi_3-\psi_3\varphi_2 \\
\psi_1\varphi_2-\psi_2\varphi_1 \\
\psi_3\varphi_1-\psi_1\varphi_3 \\
  \ea
  \right) \sim 3_A~.
  \label{tensorp}
   \ee

We formulate our model in the framework of the see-saw mechanism, even though it would also be possible to build a version (see sect. 5) where light neutrino masses are directly described by a single set of higher dimensional operators, violating the total lepton number by two units. For this we assign the 3 generations of left-handed (LH) lepton doublets $l$ and of right-handed (RH) neutrinos $\nu^c$ to two triplets $3$, while the RH charged leptons $e^c$, $\mu^c$ and $\tau^c$ all transform as $1$ (while the most usual classification in $A_4$ models is as $1$, $1''$ and $1'$). The $A_4$ symmetry is then broken by suitable flavons. All the flavon fields are singlets under the Standard Model gauge group.  The complete flavour symmetry is $A_4\times Z_4$.  We adopt a supersymmetric context, so that two Higgs doublets $h_{u,d}$, invariant under $A_4$, are present in the model. A $U(1)_R$ symmetry related to R-parity and the presence of driving fields in the flavon superpotential are common features of supersymmetric formulations. 
The field content and the symmetry assignments are as in Tab.\ref{transform}.
\begin{table} [h]
\centering
\begin{tabular}{|c||c|c|c|c|c||c|c|c|c|c|c||c|c|c|c|}
\hline
{\tt Field}& $\nu^c$ & $\ell$ & $e^c$ & $\mu^c$ & $\tau^c$ & $h_d$ & $h_u$& 
$\varphi_T$ &  $\xi'$ & $\varphi_S$ & $\xi$ & $\varphi_0^T$  & $\varphi_0^S$ & $\xi_0$\\
\hline
$A_4$ & $3$ & $3$ & $1$ & $1$ & $1$ & $1$ &$1$ &$3$ & $1'$ & $3$ & $1$ &  $3$ &  $3$ & $1$\\
\hline
$Z_4$ & -1 & i & $1$ & i & -1 & 1 & i &
i & i  & $1$ & $1$ & -1 &  $1$ & $1$\\
\hline
$U(1)_R$ & $1$& $1$ & $1$ & $1$ & $1$ & $0$ & $0$ & $0$& $0$  & $0$ & $0$ & $2$ & $2$ & $2$\\
\hline
\end{tabular}
\caption{\it Transformation properties of leptons, electroweak Higgs doublets and flavons under $A_4 \times
Z_4$ and $U(1)_R$~.}
\label{transform}
\end{table}
For the class of models of ref.\cite{lin1,lin2} the crucial feature is the alignment
\bea
\label{solS}
\langle \varphi_S \rangle =(v_S, v_S,v_S) \\ 
\langle \xi \rangle =u \nn
\eea 
\be
\langle \xi' \rangle =u' \ne 0~,~~~~~~~\langle \varphi_T \rangle =(0, v_T,0)~,~~~~~~~v_T=-\frac{h_1u'}{2h_2}~.
\label{solT}
\ee
Note that this differs from the usual $A_4$ alignment in that $\langle \varphi_T \rangle =(0, v_T,0)$ replaces $\langle \varphi_T \rangle =(v_T,0,0)$. The difference is that, while $(1,0,0)^n=(1,0,0)$ (i.e. all positive powers are aligned in the same direction), for $(0,1,0)$ we have
$(0,1,0)^2=(0,0,1)$ and $(0,1,0)^3=(1,0,0)$. These 3 directions are important in order to obtain the observed hierarchy of charged lepton masses: the electron, muon and tauon masses arise at order  $(\langle \varphi_T \rangle/\Lambda)^3$, $(\langle \varphi_T \rangle/\Lambda)^2$,
and $\langle \varphi_T \rangle/\Lambda$, respectively, where $\Lambda$ is the cutoff. For
$\langle \varphi_T \rangle/\Lambda\sim {\cal O}(\lambda_C^2)$, with $ \lambda_C$ being the Cabibbo angle, the correct hierarchy is reproduced. In the following we first assume that the stated alignment actually occurs and describe the LO structure of the model. Then in subsect. 2.3 we will show that the alignment is indeed naturally realized at LO from the most general superpotential allowed by the symmetry of the model.

\subsection{Charged leptons}

The leading order structure of the vacua in eqs.(\ref{solS},\ref{solT}) automatically generates a diagonal charged lepton matrix, through the following superpotential terms:
\bea
\label{wchl}
{w_l}&=&\frac{y_\tau}{\Lambda} \tau^c (\ell \varphi_T) \, h_d +\nn \\
&&\frac{y_\mu}{\Lambda^2} \mu^c (\ell \varphi_T \varphi_T) \, h_d +
\frac{y_\mu'}{\Lambda^2} \mu^c (\ell \varphi_T)^{''} \xi' \, h_d + \label{oplept}\\
&&\frac{y_e}{\Lambda^3} e^c (\ell \varphi_T\varphi_T)^{''} \xi' \, h_d +
\frac{y_e'}{\Lambda^3} e^c (\ell \varphi_T)' \xi^{'2} \, h_d +
\frac{y_e''}{\Lambda^3} e^c (\ell \varphi_T)' (\varphi_T \varphi_T )'' \, h_d +\nn \\
&&\frac{y_e'''}{\Lambda^3} e^c (\ell \varphi_T)'' (\varphi_T \varphi_T )' \, h_d +
\frac{y_e^{\rm iv}}{\Lambda^3} e^c (\ell \varphi_T)_1 (\varphi_T \varphi_T )_1 \, h_d+..... \nn
\eea
In the above expression for the superpotential $w_l$, for each charged lepton flavour, only the lowest order operators
in an expansion in powers of $1/\Lambda$ are explicitly shown. Dots stand for higher
dimensional operators that will be discussed later on. Note that the $Z_4$ parities impose different powers of $\varphi_T$ and/or $\xi'$ for the electron, muon and tauon terms, while $\varphi_S$ and $\xi$ are invariant under $Z_4$ and only appear in non leading terms as additional factors. After symmetry breaking, the mass matrix has the form:
\be
m_\ell=\left(
\begin{array}{ccc}\frac{v_T v_d}{\Lambda^3}\left(2 y_e v_T u'+y_e' u'^2 + y_e'' v_T^2\right)& 0& 0\\
0& \frac{v_T v_d}{\Lambda^2}\left(2 y_\mu v_T+y_\mu' u'\right)& 0\\
0& 0& \frac{y_\tau v_d v_T}{\Lambda}
\end{array}
\right)~~~,
\label{mlept}
\ee
where $v_d=\langle h_d \rangle$. Note that the operators in the last line of eq. (\ref{wchl}) vanish when the LO  VEV's given in eqs.(\ref{solS},\ref{solT}) are inserted. 

As a result, the charged lepton mass matrix is diagonal and with hierarchical entries. To estimate the order of magnitude of $v_T$ and $u'$, we can use the experimental information on the ratio of lepton masses. Assuming that all the $y$ coefficients are of ${\cal O}(1)$, one obtains: 
\bea
\left(\frac{m_\mu}{m_\tau}\right)&\sim&2\,\varepsilon+\varepsilon_u \simeq 0.06 \nn \\
\left(\frac{m_e}{m_\tau}\right)&\sim&2\,\varepsilon\,\varepsilon_u+\varepsilon^2 +\varepsilon^2_u \simeq 0.0003 \nn 
\eea
where we introduced the small quantities 
\bea
\nn
\varepsilon=v_T/\Lambda \qquad \varepsilon_u=u'/\Lambda.
\eea
These relations are satisfied for both sets of values:
\bea
(\varepsilon,\varepsilon_u)&=&(0.043,-0.025) \qquad (\varepsilon,\varepsilon_u)=(0.077,-0.094).
\eea
As we see, we can roughly assume that both $\varepsilon$ and $\varepsilon_u$ are of the same order of magnitude, ${\cal O}(\lambda_C^2)$ (see also eq.(\ref{eps})). With these assumptions, and using the relation connecting $u'$ to $v_T$ of eq.(\ref{solT}) we can rewrite the lepton matrix in a simplified form:
\be
m_\ell=\varepsilon\,v_d\left(
\begin{array}{ccc}\varepsilon^2\left[-4\frac{h_2}{h_1} (y_e -\frac{h_2}{h_1} \, y_e')  + y_e'' \right]& 0& 0\\
0& 2\,\varepsilon\,(y_\mu -\frac{h_2}{h_1} \,y_\mu' )& 0\\
0& 0& y_\tau 
\end{array}
\right).
\label{mleptsimpl}
\ee

\subsection{Neutrinos}

In the neutrino sector the superpotential is given by:
\bea
\label{lagneu}
w_\nu = y_\nu (\nu^c \ell)\,h_u + (M+a\,\xi)\,\nu^c\nu^c+b \,\nu^c\nu^c\,\varphi_S
\eea
where $a$ and $b$ are generic coefficients and $M$ is a constant with dimension of mass. Note that we also included in the LO neutrino superpotential linear terms in
$\varphi_S$ and $\xi$. 

The Dirac mass matrix is obtained from the first term in eq.(\ref{lagneu}) and it is given by:
\bea
\label{mdir}
m_D=y_\nu\, v_u\left(
\begin{array}{ccc}
1  &  0   & 0 \\
0 & 0  &  1 \\
 0& 1  &  0
\end{array}
\right) = v_u\, Y_\nu \ .
\eea
The other terms lead to the Majorana mass matrix:
\bea
\label{mmleading}
m_M=\left(
\begin{array}{ccc}
M+a\,u+2\,b\,v_S  &  -b\,v_S   & -b\,v_S \\
-b\,v_S & 2\,b\,v_S  &  M+a\,u-b\,v_S \\
-b\,v_S & M+a\,u-b\,v_S  &  2\,b\,v_S
\end{array}
\right)
\eea
with eigenvalues:
\bea
\label{majei}
M_1&=&M + a \,u + 3\, b\, v_S \nn \\
M_2&=&M + a \,u  \\
M_3&=&-M - a \,u +3\, b\, v_S ~.\nn
\eea
The light neutrino mass matrix is then given by the see-saw formula
\bea
m_{light}= -m_D^T\, m_M^{-1}\, m_D~.
\eea
Note that all matrices $m_D$, $m_M$, $m_M^{-1}$ and $m_{light}$ are of the general form
\bea
\label{massgen}
m=\left(
\begin{array}{ccc}
x  &  y   & y \\
y & x+v  &  y-v \\
y & y-v  &  x+v
\end{array}
\right) \ 
\eea
and therefore are diagonalized by $U_{\rm TB}$  (see, for example,\cite{GALT}): 
\be
U_{\rm TB}=\left(
\begin{array}{ccc}
\sqrt{2/3}& 1/\sqrt{3}& 0\\
-1/\sqrt{6}& 1/\sqrt{3}& -1/\sqrt{2}\\
-1/\sqrt{6}& 1/\sqrt{3}& +1/\sqrt{2}
\end{array}
\right).
\label{HPSmatrix1}
\ee
The $m_{light}$ eigenvalues are:
\bea
m_1&=&-\,\left(\frac{v_u^2\,y_\nu^2}{M + a \,u + 3\, b\, v_S} \right) \nn \\
m_2&=&-\,\left(\frac{v_u^2\,y_\nu^2}{M + a \,u} \right) \\
m_3&=& \left(\frac{v_u^2\,y_\nu^2}{M + a \,u - 3\, b\, v_S} \right).\nn
\eea

\subsection{Alignment}

At LO the most general driving superpotential $w_d$ invariant under $A_4 \times Z_4$ with $R=2$ is given by
\bea
\label{phis}
w_d&=&M (\varphi_0^S \varphi_S)+g_1 (\varphi_0^S \varphi_S\varphi_S)+
g_2 \xi (\varphi_0^S \varphi_S)+
g_3 \xi_0 (\varphi_S\varphi_S)+
g_4 \xi_0 \xi^2+
M_\xi \xi_0 \xi  \\ \nn
 &+&  M_0^2 \, \xi_0 +
h_1 \xi' (\varphi_0^T \varphi_T)''+
h_2 (\varphi_0^T \varphi_T\varphi_T)~.\label{wd2}
\eea

We now discuss the vacuum alignment configuration determined by $w_d$.
The equations giving the vacuum structure for the fields $\varphi_T$ and $\xi'$ are:
\bea
\frac{\partial w}{\partial \varphi^T_{01}}&=&2 h_2({\varphi_T}^2_1-{\varphi_T}_2\,{\varphi_T}_3)+
h_1\,\xi'\,{\varphi_T}_3=0\nn\\
\frac{\partial w}{\partial \varphi^T_{02}}&=&2 h_2({\varphi_T}^2_2-{\varphi_T}_1\,{\varphi_T}_3)+
h_1\,\xi'\,{\varphi_T}_2=0 \label{sys:t} \\
\frac{\partial w}{\partial \varphi^T_{01}}&=&2 h_2({\varphi_T}^2_3-{\varphi_T}_1\,{\varphi_T}_2)+
h_1\,\xi'\,{\varphi_T}_1=0\nn
\eea
whose solutions are:
\be
\langle \xi' \rangle =u' \ne 0~,~~~~~~~\langle \varphi_T \rangle =(0, v_T,0)~,~~~~~~~v_T=-\frac{h_1u'}{2h_2}~.
\ee
with $u'$ undetermined.

From Eq.~(\ref{phis}), we can obtain the equations from which to extract the vacuum expectation values for 
$\varphi_S$ and $\xi$:
\bea
\frac{\partial w}{\partial \varphi^S_{01}}&=&2 g_1({\varphi_S}^2_1-{\varphi_S}_2\,{\varphi_S}_3)+
(M+g_2\,\xi)\,{\varphi_S}_1=0\nn\\
\frac{\partial w}{\partial \varphi^S_{02}}&=&2 g_1({\varphi_S}^2_2-{\varphi_S}_1\,{\varphi_S}_3)+
(M+g_2\,\xi)\,{\varphi_S}_3=0 \label{sys:s} \\
\frac{\partial w}{\partial \varphi^S_{03}}&=&2 g_1({\varphi_S}^2_3-{\varphi_S}_1\,{\varphi_S}_2)+
(M+g_2\,\xi)\,{\varphi_S}_2=0\nn \\
\frac{\partial w}{\partial \xi_0}&=&
M_\xi \xi+g_3({\varphi_S}^2_1+2{\varphi_S}_2\,{\varphi_S}_3)+g_4\xi^2 + M_0^2=0 ~.\nn
\eea

An extremum solution is given by: 
\bea
\langle \varphi_S \rangle =(v_S, v_S,v_S) \\ 
\langle \xi \rangle =u \nn
\eea 
with the conditions:
\bea
u&=&-\frac{M}{g_2} \\
v_S^2&=&  \frac{1}{3 g_2^2 g_3} \left[g_2 (M M_\xi -g_2 M_0^2)-g_4 M^2\right].
\eea
Note that we expect a common order of magnitude for the VEV's (scaled by the cutoff $\Lambda$):
\be
\frac{v_T}{\Lambda}\sim \frac{u'}{\Lambda}\sim \varepsilon,~~~~~~~\frac{v_S}{\Lambda}\sim \frac{u}{\Lambda}\sim \varepsilon'~. \\
\label{eps}
\ee
However, as the minimization equations for the two sets are separate, we can tolerate a moderate hierarchy between $\varepsilon$ and $\varepsilon'$.

It is easy to check that the above solutions for the vacuum expectation values correspond to 
isolated directions, in the sense that if one perturbs the vacua in eqs.(\ref{solS},\ref{solT}) by small quantities (that is $\langle \varphi_T \rangle=(\varepsilon_1, v_T+\varepsilon_2,\varepsilon_3)$ and $\langle \varphi_S \rangle =(v_S+\varepsilon_1^s, v_S+\varepsilon_2^s,v_S+\varepsilon_3^s)$) the minimizing equations do not allow to continuously shift the given directions. In fact it is important to note that the alignment direction for $\varphi_S$  corresponds to a particular subgroup $G_S$ of  $A_4\times Z_4$ generated by the operator $S$ which is left invariant. On the contrary $\varphi_T$ is aligned along a direction which breaks  $A_4\times Z_4$ completely. It corresponds however to a subgroup of $A_4\times Z_3$ which is an accidental symmetry of the relevant terms in the LO superpotential \cite{lin1} (those involved in the minimization that leads to the $\varphi_T$ VEV). This accidental symmetry does not survive at the next to the leading order level, where in fact the alignment is displaced.

We conclude that at the LO the present model leads to diagonal and hierarchical charged leptons and to exact TB mixing for neutrinos.

\section{Beyond the leading order}

At the next level of approximation each term $w_l$, $w_\nu$ and $w_d$ of the superpotential is corrected by operators of higher dimension whose contributions are suppressed by at least one power of VEV's/$\Lambda$.  The corrections to $w_d$ determine small deviations from the LO VEV alignment configuration. The next to the leading order (NLO) corrections to mass and mixing matrices are obtained by inserting the corrected VEV alignment in the LO operators plus the contribution of the new operators evaluated with the unperturbed VEV's. A crucial feature is that $\varphi_S$ and $\xi$ are invariant under $Z_4$ so that it is always possible to include an extra power of these fields on top of each LO term (the fact that $\varphi_S$ is a 3 under $A_4$ can be taken into account by a suitable reshuffling of the $A_4$ contractions in order to form an invariant). On the contrary $\varphi_T$ and $\xi'$ take a phase i under $Z_4$ so that  adding such an extra factor to a given LO term is not possible. The results, which follow from a detailed analysis, can be summarized as follows. 

The VEV configuration obtained from $w_d+\Delta w_d$, where $\Delta w_d$ is the most general set of terms suppressed by one power of the cutoff is given by:
\bea
\langle \varphi_S \rangle &=&(v_S+\delta v_S , v_S+\delta v_S ,v_S+\delta v_S ) \nn \\
\langle \varphi_T \rangle &=&(\delta v_{T_1}, v_T+\delta v_{T_2}, \delta v_{T_3}) \label{vaccorr}\\
\langle \xi \rangle &=& u+\delta u \nn 
\eea
and $u'$ is still undetermined. Thus $\langle \varphi_S \rangle$ acquires ${\cal O}(1/\Lambda$) corrections in the same direction, whereas all components of $\langle \varphi_T \rangle$ acquire different corrections so that its alignment is tilted. 

In the charged lepton sector, the correction $\Delta w_l$ is obtained by adding to each term of $w_l$ one factor of
$\varphi_S/\Lambda$ or $\xi/\Lambda$  in all possible ways with arbitrary coefficients. As a result, each diagonal entry gets a small correction, while all non diagonal entries become non vanishing and of the order of the diagonal term in each row multiplied by $\varepsilon'$:
\be
m_\ell=\varepsilon\,v_d\left(
\begin{array}{ccc} a_1 \, \varepsilon^2 &a_2 \, \varepsilon^2\, \varepsilon'& a_3 \,\varepsilon^2 \, \varepsilon'\\
b_1\, \varepsilon\, \varepsilon'& b_2\,\varepsilon & b_3\,\varepsilon\, \varepsilon'\\
c_1\,\varepsilon' & c_2\,\varepsilon' & c_3
\end{array}
\right) 
\label{mleptfin}
\ee
where the coefficients $a_i, b_i$ and $c_i$ are ${\cal O}(1)$ unspecified constants. This pattern is not altered when one adds the corrections from inserting the shifted VEV's in the LO expression of $w_l$. In fact the shifts $\delta v_i/\Lambda$ are also of order $\varepsilon'$ and the corresponding corrective terms contain one additional power of $\delta v_i/\Lambda$ for each matrix element in a row which contains a fixed power of $\varepsilon$ to start with. Thus, including these additional corrections only amounts to a redefinition of the $a_i, b_i$ and $c_i$ coefficients. 

The matrix 
$m_\ell^\dagger\, m_\ell$ can be diagonalized to 
${\rm Diag}[|a_1^2 \, \varepsilon^4|,|b_2^2\,\varepsilon^2|,|c_3^2|] $ by the unitary transformation 
\be
U_\ell= \left(
\begin{array}{ccc} 1 & (\frac{b_1}{b_2}\,\varepsilon')^* &  (\frac{c_1}{c_3} \,\varepsilon')^*  \\
-\frac{b_1}{b_2}\,\varepsilon'& 1 & (\frac{c_2}{c_3}\, \varepsilon')^*\\
-\frac{c_1}{c_3} \,\varepsilon'& -\frac{c_2}{c_3} \,\varepsilon' & 1
\end{array}
\right)~. 
\label{ue}
\ee
Note that, at this order, the coefficients of the electron row in eq.(\ref{mleptfin})
do not enter in $U_\ell$. From the matrix in eq.(\ref{ue}), we can compute the corresponding corrections to the TB mixing matrix according to $U_{\rm PMNS}=U_\ell^\dagger\, U_{\rm TB}$
and all entries of $U_{\rm TB}$ get corrected to ${\cal O}(1/\Lambda)$.

In the neutrino sector the corrections due to inserting the VEV shifts in the LO operators do not affect the Dirac mass at all while the changes in the Majorana mass are still of the form that is diagonalized by the TB mixing matrix. However the corrections from operators of higher dimension obtained by inserting one extra power of $\varphi_S/\Lambda$ or $\xi/\Lambda$, of order $\varepsilon'$, alter both the Dirac and the Majorana mass in such a way that the TB mixing pattern is completely violated by small corrective terms. To be precise the Dirac neutrino mass directly receives a relative correction of 
${\cal O}(\varepsilon')$. The Majorana neutrino mass in LO has terms of order $M$ and of order $v_S$ (see eq.(\ref{lagneu})). At NLO it receives TB mixing violating corrections of order $v_S \varepsilon'$.
As a result, the overall correction to TB mixing arises from the most general symmetric matrix of order $v_S \varepsilon'$:
\bea
m_\nu=-m_D^T\, m_M^{-1}\, m_D= (m_\nu)_{TB}+v_S \varepsilon' 
\left(\begin{array}{ccc}
 A&  B     &   C   \\
  B  & D    &  F \\
C   & F    &  E   
\end{array}\right)
\eea
where by $(m_\nu)_{TB}$ we denote the matrix, diagonalizable by $U_{TB}$, which is obtained from the LO term plus the corrective terms that can be reabsorbed in the LO coefficients as they preserve TB mixing.

In conclusion when the NLO corrections are included TB mixing is violated by small terms and one expects:
\bea
\sin^2\theta_{12}=\frac{1}{3}+{\cal O}(\varepsilon')\nn \\
\sin^2\theta_{23}=\frac{1}{2}+{\cal O}(\varepsilon') \label{corr}\\
\sin\theta_{13}={\cal O}(\varepsilon')~.\nn 
\eea
We have already noted that the data require that $\varepsilon' \lesssim {\cal O}(\lambda_C^2)$.

\section{Light neutrino spectrum and constraints from $r$}
 
It is useful to study the constraints on the model imposed by the observed values of $\Delta m^2_{atm}$ and of the ratio $r=\Delta m^2_{sol}/|\Delta m^2_{atm}|$. Here $\Delta m^2_{sol}=|m_2|^2-|m_1|^2 > 0$, $\Delta m^2_{atm}=|m_3|^2-|m_1|^2$. We do this in the LO approximation. In fact the results of the previous section indicate that the corrections to the spectrum are sufficiently small to be neglected for a first orientation. For this discussion we adopt the following parameterization:
\bea
A &=& M+a\,u = |A|\,e^{i\,\phi_A}\nn \\
B&=&3\,b\,v_S= |B|\,e^{i\,\phi_B}\nn \\
\alpha&=&\frac{|B|}{|A|} \label{param} \\
\phi&=&\phi_B-\phi_A ~.\nn 
\eea
In LO the masses can then be written as:
\bea
m_1&=&-\frac{v_u^2\,y_\nu^2}{|A|\,e^{i\,\phi_A}}\,\left(\frac{1}{1+\alpha \, e^{i\,\phi}}\right) \nn \\
m_2&=&-\frac{v_u^2\,y_\nu^2}{|A|\,e^{i\,\phi_A}}\label{emmei} \\
m_3&=&\frac{v_u^2\,y_\nu^2}{|A|\,e^{i\,\phi_A}}\,\left(\frac{1}{1-\alpha \, e^{i\,\phi}}\right)~.\nn
\eea
Note that, in the leading approximation, the model predicts the relation:
\be
\frac{1}{m_3}=\frac{1}{m_1}-\frac{2}{m_2}~.\\
\ee
This LO relation is typical of see-saw versions of $A_4$. Here it is expected to hold up to corrections
of $O(\lambda_C^2)$, as discussed in section 3. Note 
that $m_i$ $(i=1,2,3)$
are complex numbers, so that the previous relation cannot
be used to exactly predict one physical neutrino mass in terms of the other 
two ones, as is clear by the following discussion. Nevertheless, it provides a non-trivial constraint that the neutrino masses should obey.

In terms of the parameters $\alpha$ (which is real and positive) and $\phi$, the ratio $r$ is:
\bea
\label{eq:erre}
r&=&\frac{\Delta m^2_{sol}}{|\Delta m^2_{atm}|}=\frac{(1+\alpha^2-2\,\alpha\,\cos \phi)(\alpha+2\cos \phi)}{4\,|\cos \phi|}~.
\eea
The limit $\alpha \to 0$ gives $r=1/2$, which is too large compared to the experimental value.  
Thus in order to accomodate $r \sim 1/30$ one needs a value of $\alpha$ of ${\cal O}(1)$, which implies that (see eq.(\ref{param})) $|A| = |M+a\,u| \sim |B|=3\,|b\,v_S|$. With $|a|, |b| \sim {\cal O}(1)$ this is obtained if $|M|  \sim u, v_S$, or, in other words, $|M|$ must be sizeably smaller than the cutoff $\Lambda$. We interpret this result as related to the fact that the RH neutrino Majorana mass $M$ must empirically be close to $M_{GUT}$.  This means that in the context of a grand unified theory $M$ must be of $\mathcal{O}(M_{GUT})$ rather than of $\mathcal{O}(M_{Planck})$.
More precisely, from Eq.~(\ref{eq:erre}) one recognizes that a small r can be reproduced  if 
\bea
a) ~\cos \phi &\sim& \alpha \sim 1 \nn \\
\label{cond}
\\
b)~\cos \phi &=& -\frac{\alpha}{2} +\delta \alpha \qquad \delta \alpha \sim {\cal O}(r)~. \nn
\eea
The first condition $a)$ corresponds to a normal hierarchy spectrum whereas the second condition $b)$ leads to an inverted hierarchy scheme.  

In the normal hierarchy case, $a)$, by expanding around $\cos \phi \sim \alpha \sim 1$, we obtain:
\bea
|m_1|^2&=&\frac{1}{3}
\Delta m^2_{atm}~r+...\nn\\
|m_2|^2&=&\frac{4}{3}
\Delta m^2_{atm}~r+...\nn\\
|m_3|^2&=&\left(1+\frac{r}{3}\right)
\Delta m^2_{atm}+...\nn\\
|m_{ee}|^2&=&\frac{16 }{27}\Delta m^2_{atm}~r+...\sim (0.007~\rm{eV})^2~~~,
\label{lospe2}
\eea
where we have expressed the parameters in terms of $\Delta m^2_{atm}$ and $r$. Dots denote terms of order $r^2$ as well as corrections beyond the LO. Note that in this model $|m_1|$ cannot vanish. In the last line $|m_{ee}|$ is the effective mass combination controlling the violation of the total lepton number in neutrinoless double beta decay. 

\begin{figure}[h!]
\centering
\hspace{-0.3cm}
\includegraphics[width=0.5\linewidth]{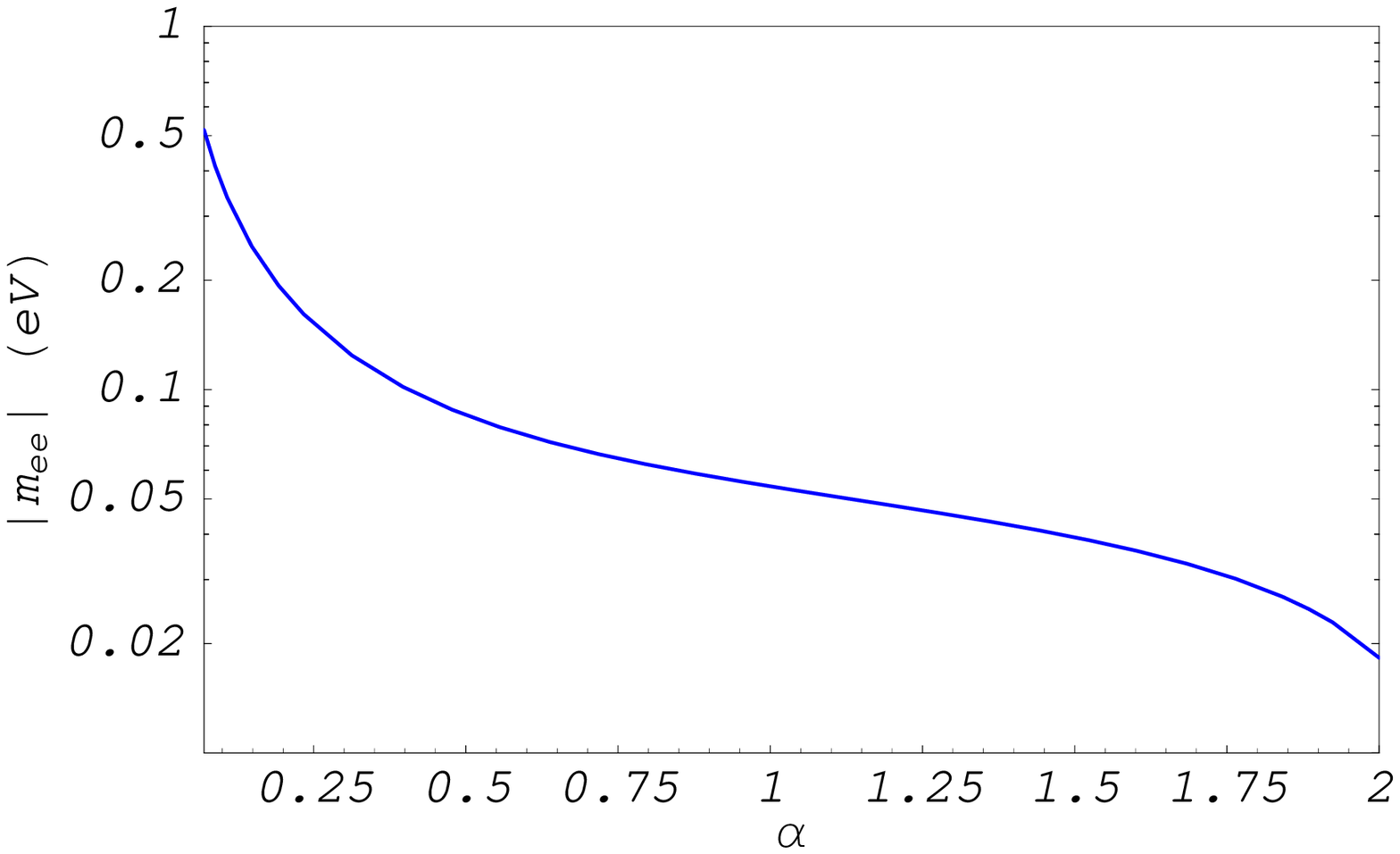} 
\includegraphics[width=0.5\linewidth]{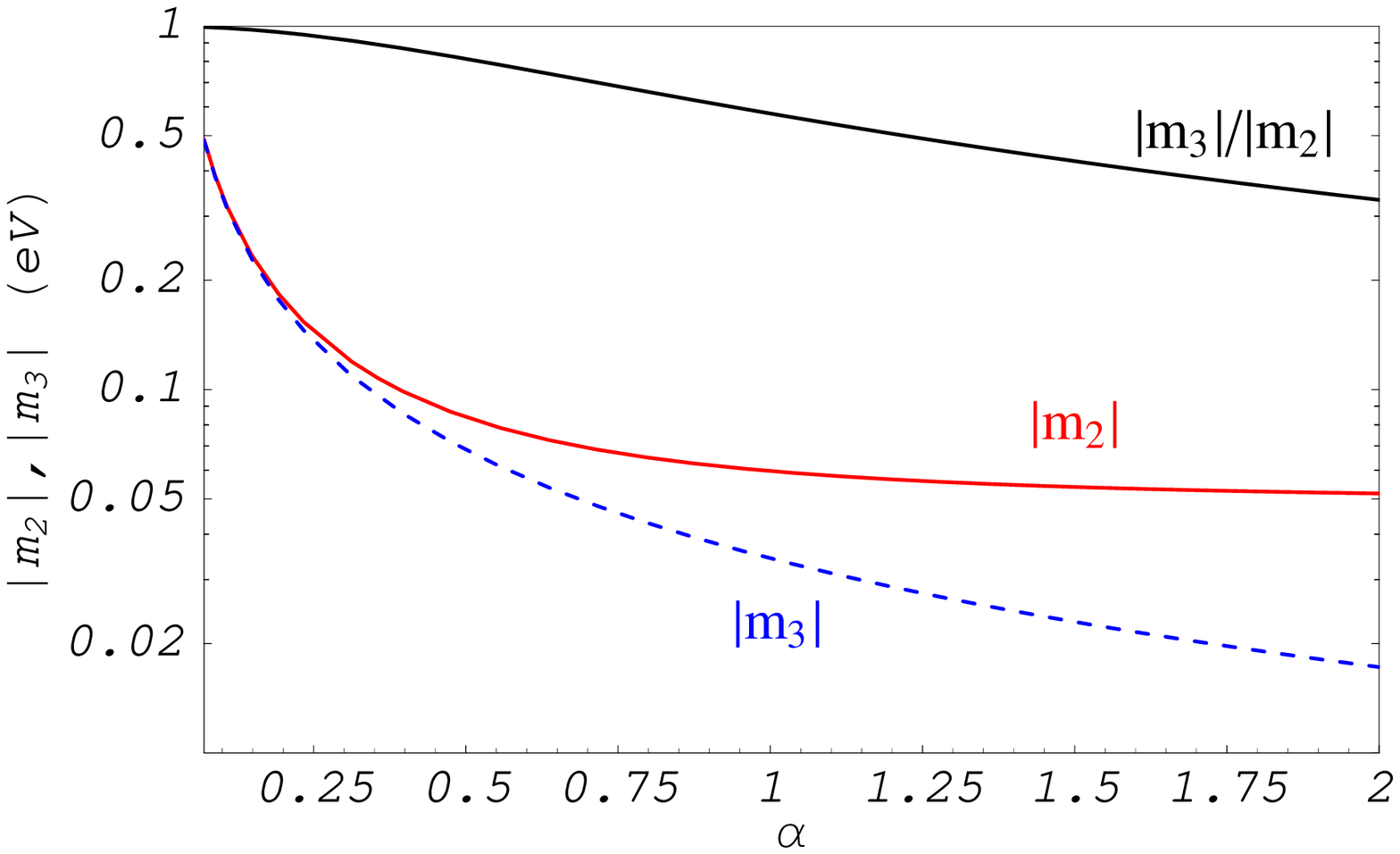}
\caption{\label{fig:meemasses} \it Behaviour of neutrino masses  in the inverted hierarchy case (at fixed $\Delta m^2_{atm}$ and $r$) as a function of $\alpha$ in the range between 0.07 and 2 (the lower bound on $\alpha$ corresponds to an upper bound on $|m_i|$). Left panel: $|m_{ee}|$. Right panel: $|m_2|$, $|m_3|$ and the ratio $|m_3|/|m_2|$.}
\end{figure}

In the inverse hierarchy case $b)$, we set $2\cos\phi=-\alpha+\delta$ with $\delta$ positive and small and vary $\alpha$ in the range between 0.07 and 2 (the lower limit comes from the absolute bound on the squared masses at fixed $\Delta m^2_{atm}$ and $r$, taken indicatively at $|m_i| \lesssim 0.5~eV$). The quantity $\delta$ is determined in terms of $r$:
\beq
\delta=\frac{2\alpha r}{1+2\alpha^2}\\
\label{delta}
\eeq
while the scale of the squared masses is fixed by $|\Delta m^2_{atm}|$. In terms of $\alpha$ one obtains\footnote{Note that earlier results found in the literature like, e.g., in  \cite{AFmodular} and \cite{AFH},
were obtained from a different type of expansion and for this reason the linear terms in $r$ are not the same.}:
\bea
|m_2|^2&=&|m_1|^2+|\Delta m^2_{atm}|\,r=
|\Delta m^2_{atm}|\left(\frac{1+2\alpha^2}{2\alpha^2}\right)\left[1+r\left(1+\frac{1}{(1+2\alpha^2)^2}\right)\right]+...\nn\\
|m_3|^2&=&|m_1|^2-|\Delta m^2_{atm}|\,.\label{lospe}
\eea
Note that the $r$ terms are sufficiently small that the corresponding contributions could be overshadowed by the neglected NLO corrections. Omitting these additional linear terms in $r$, $m_{ee}$ is given by:
\beq
|m_{ee}|^2=|\Delta m^2_{atm}| \left(\frac{1+2\alpha^2 }{2\alpha^2}\right)\left(1-\frac{2}{9}\alpha^2\right)+...\\
\label{ee}
\eeq
$|m_{ee}|^2$ is a decreasing function of $\alpha$ in the physical range and is close to $|\Delta m^2_{atm}|\sim (0.05~\rm{eV})^2$ for $\alpha=1$ and to $|\Delta m^2_{atm}|/8$ for $\alpha=2$. The behaviour of 
$|m_{ee}|$ as a function of $\alpha$, including the neglected terms in $r$, is shown in the left panel of 
Fig.(\ref{fig:meemasses}).
The most pronounced inverse hierarchy is realized for $\alpha=2$ where $|m_3| \sim |m_1|/3$, as it is seen from
the right panel of Fig.(\ref{fig:meemasses}). 

\begin{figure}[h!]
\centering
\hspace{-0.3cm}
\includegraphics[width=0.5\linewidth]{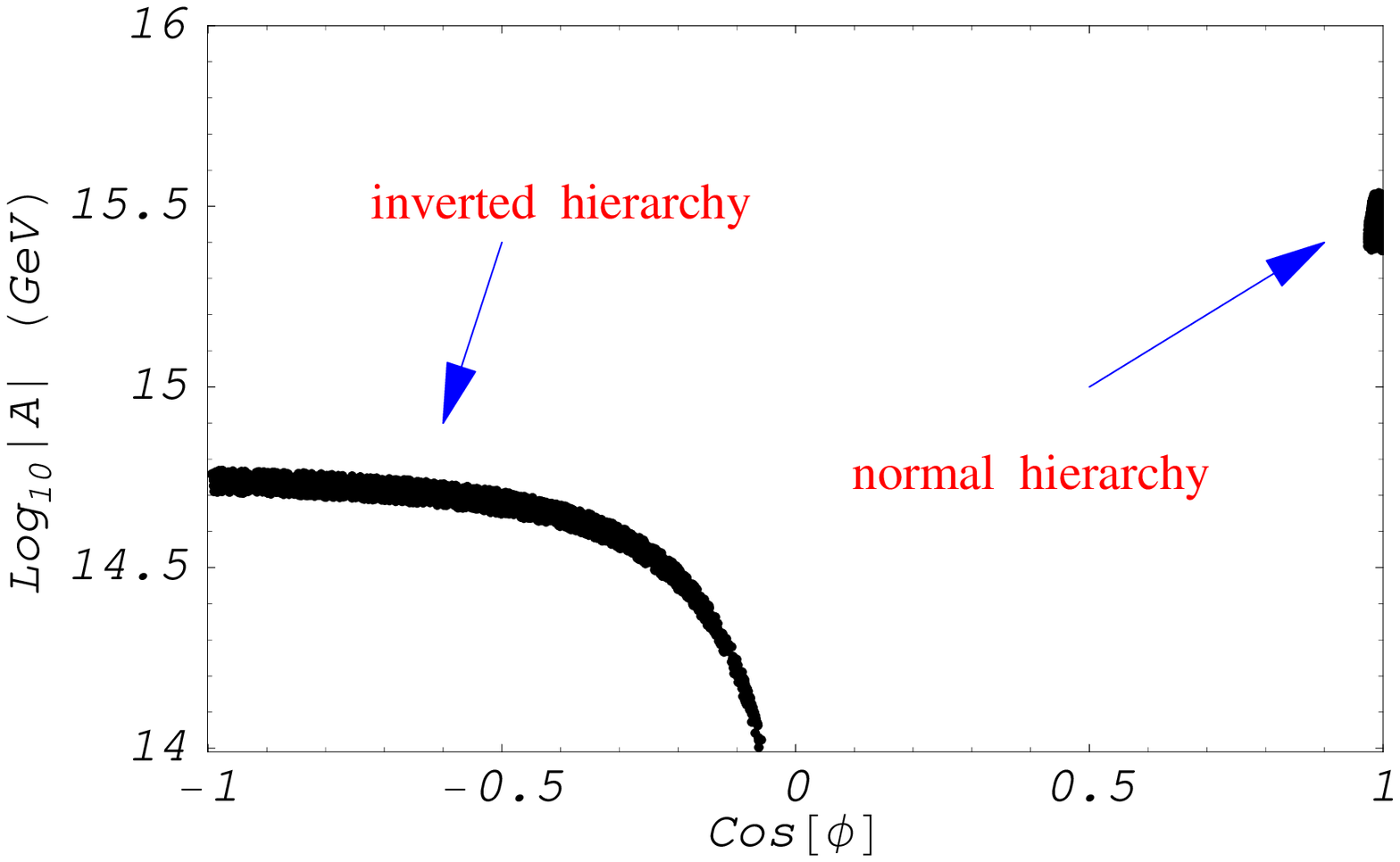} 
\includegraphics[width=0.5\linewidth]{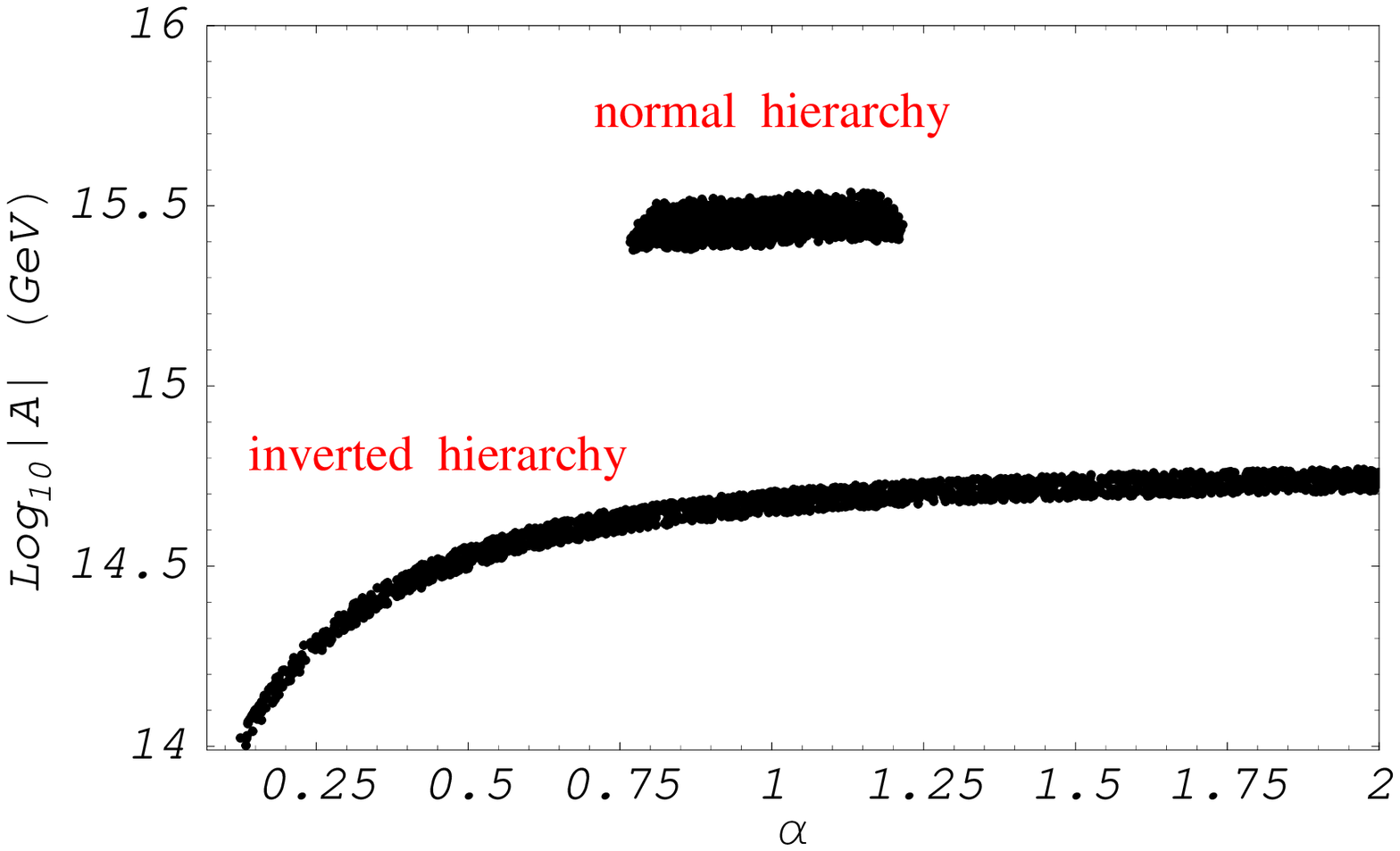}
\caption{\label{fig:M} \it Values of $|A|$ (in log-scale) as a function of $\cos \phi$ (left panel) and $\alpha$ (right panel) obtained by imposing  the relations in eq.(\ref{condnorm}).}
\end{figure}

Using reasonable values for the parameters, like $v_u=170$ GeV and $y_\nu=1$, 
we estimate $|A|\sim {\cal O}(10^{14}-10^{15})$ GeV (see eq. (\ref{param})), with a significant preference of the normal (inverted) hierarchy toward larger (smaller) values of $|A|$,
as we can see in Fig.(\ref{fig:M}), where we show the behaviour of $|A|$ as a function of $\cos \phi$ (left panel) and  $\alpha$ (right panel) for both neutrino spectra.
The allowed parameter space is built from the 
following 3-$\sigma$ constraints \cite{MaltoniIndication}:
\bea
\Delta m^2_{sol} &>& 0 \nn \\
|\Delta m^2_{atm}| &=& \pm 2.41 \pm 0.34\times 10^{-3}\, eV^2 \label{condnorm} \\
r&=& 0.032 \pm 0.006\nn~. 
\eea

In conclusion the observed value of $r$ requires that $M$ is at or just below $M_{GUT}$. Once this is realized the observed spectrum can be reproduced.

\section{Effective operators}
In addition to the previous operators, we should also consider those higher dimensional operators that could arise beyond the type-1 see-saw terms. The following terms contribute to the effective light neutrino mass operators:

\bea
W_\nu^{\rm eff} &=& \frac{C}{\Lambda} \left(\ell \, h_u \,\ell \, h_u\right)
+\frac{D}{\Lambda^2}  \left(\ell \, h_u \,\ell \, h_u\right)\,\varphi_S + 
 \frac{E}{\Lambda^2} \left(\ell \, h_u \,\ell \, h_u\right)\,\xi \nn \\
 &=& \frac{C'}{\Lambda} \left(\ell \, h_u \,\ell \, h_u\right)
+\frac{D}{\Lambda^2}  \left(\ell \, h_u \,\ell \, h_u\right)\,\varphi_S~.
\eea
The redefinition in the last line is justified because the operator containing $\xi$ gives the same mass matrix as the leading order operator proportional to $C$.
This results in the following matrix:
\bea
\label{meffleading}
m_{eff}=\frac{v_u^2}{\Lambda}\left(
\begin{array}{ccc}
C'+ 2 D\,\frac{ v_S}{\Lambda}  &  -D\,\frac{ v_S}{\Lambda}   & -D\,\frac{ v_S}{\Lambda} \\
-D\,\frac{ v_S}{\Lambda} & 2 D\,\frac{ v_S}{\Lambda} &  C'-D\,\frac{ v_S}{\Lambda} \\
-D\,\frac{ v_S}{\Lambda}' & C'-D\,\frac{ v_S}{\Lambda}  &  2 D\,\frac{ v_S}{\Lambda}
\end{array}
\right)
\eea
which is diagonalized by $U_{TB}$. Its eigenvalues contribute to the light neutrino mass matrix as follows:
\bea
m_1&=&v_u^2\,\left(\frac{-y_\nu^2}{M + a \,u + 3\, b\, v_S}+\frac{C'}{\Lambda}+3\,D\,\frac{v_S}{\Lambda^2}\right) \nn \\
m_2&=&v_u^2\,\left(\frac{-y_\nu^2}{M + a \,u}+\frac{C'}{\Lambda}\right) \label{efo}\\
m_3&=&v_u^2\,\left(\frac{y_\nu^2}{M + a \,u - 3\, b\, v_S}-\frac{C'}{\Lambda}+3\,D\,\frac{v_S}{\Lambda^2}\right).\nn
\eea
It is natural to assume that $C'\sim D \sim {\cal O}(1)$. Then the $D$ term is subleading and the importance of the $C'$ term depend on the relative size of $M$ and $\Lambda$. We have seen in the previous section that, in order to reproduce the observed value of $r$, $M$ must be considerably smaller than $\Lambda$. Then the contribution of the effective operators in eq.(\ref{efo}) is suppressed at a level comparable to NLO corrections.

\section{Additional phenomenological consequences}

\subsection{RGE evolution of the mixing parameters}
Only for a nearly degenerate neutrino mass spectrum, the running of the angle $\theta_{12}$ from $M_{GUT}\sim 10^{15}$ GeV down to SUSY breaking scale ($M_S\sim 10^{3}$ GeV) is non negligible. The evolution is large for $|m_1|$ sufficiently close to $|m_2|$ (the two are kept apart by the non vanishing value of $r$). In our model, neglecting $r$:
\beq
\left|\frac{m_2}{m_1}\right|^2=1+\alpha^2+2\alpha \cos\phi~.\\
\label{rat}
\eeq
The ratio approaches 1 for $\alpha \sim 0$ or $\alpha+2\cos \phi \sim 0$, which means that this is the case in the whole inverse hierarchy interval.

In the case of CP conserved  neutrino mixing matrix and in the limit of vanishing electron and muon Yukawa couplings, 
the evolution equation of $\theta_{12}$ is dictated by \cite{Chankowski:2001mx}:
\bea
\label{eq:s12}
\dot s_{12} &=&\frac{1}{2}\,s_{12}\, (1-s^2_{12})\,A_{21}\,y_\tau^2
\eea 
where $s_{12}=\sin \theta_{12}$ and
\bea
A_{21} &=& \frac{|m_1+m_2|^2}{\Delta m^2_{sol}}~.
\eea 
In our model $m_1$ and $m_2$ are complex numbers and, by denoting the 2-1 relative phase as $\phi_2-\phi_1=2\beta$, the following symmetric result holds \cite{Kuo:2001ha}:
\beq
A_{21} =  \frac{|m_1+m_2|^2}{\Delta m^2_{sol}}= \cos^2\beta \frac{|m_1|+|m_2|}{|m_1|-|m_2|}+\sin^2\beta \frac{|m_2|-|m_1|}{|m_1|+|m_2|}\\
\label{Agen}
\eeq
which shows that for a generic phase an interpolation between the results for $\pm1$ phases applies.
The derivative is done with respect to the variable $t=\left(1/16\,\pi^2\right)\,\log \frac{M_{GUT}}{M}$. 
The solution of eq.(\ref{eq:s12}) can then be expressed in terms of the small quantity $\lambda=1/3-s^2_{12}$:
\bea
\lambda&=&\frac{2}{9}\, A_{21}\,\frac{y_\tau^2}{16 \pi^2}\log \frac{M_{GUT}}{M_S}~.
\eea 
Since  $\lambda \lesssim 0.05$, and given that $y_\tau^2\sim 10^{-4} (1+\tan^2\beta)$, we can avoid a running of $\theta_{12}$ beyond the experimental limits if 
\bea
\label{eq:rel}
A_{21}\,(1+\tan^2\beta) \lesssim 10^4. 
\eea

In the normal hierarchy case, $m_2\sim 2 m_1$ and both are real numbers, so that $A_{21}\sim 3$ and there is no appreciable evolution. In the inverse hierarchy case from eq.(\ref{Agen}) one obtains the following approximate expression in the physical range of $\alpha$: 
\beq
A_{21}  = \frac{|m_1+m_2|^2}{\Delta m^2_{sol}}\sim \frac{1+2\alpha^2}{2\alpha^2 r}\left(4-\alpha^2+\frac{4\alpha^2 r}{1+2\alpha^2}\right)~.\\
\label{AIH}
\eeq
At small $\alpha$ the light neutrino masses become nearly degenerate (see Fig.(\ref{fig:meemasses}) ) and we expect evolution to become important. As we have already mentioned, the existing upper limits on neutrino masses require $\alpha > 0.07$. For values of $\alpha$ near its bound, from the previous equations, we obtain $A_{12}\sim 10^4$ and the evolution is not negligible. But $A_{12}$ decreases fast with $\alpha$ and in the central region where $\alpha\sim 1$ we have $A_{12} \sim 150$, so that evolution can be neglected for $\tan\beta < 10$. Above $\alpha \sim 1$ all realistic values of $\tan \beta$ become gradually allowed. Near $\alpha = 2$, where the pattern of inverse hierarchy is most pronounced, $A_{12}=2$ and evolution effects are completely negligible.

\subsection{Leptogenesis}

In the early universe out-of-equilibrium decays  of heavy neutrinos to lepton and Higgs doublets produce lepton number asymmetries.
The asymmetry parameters are defined as follows:
\bea
\label{epslepto}
\epsilon_i&=&\frac{1}{8\pi (\hat Y \hat Y^\dagger)_{ii}}\,\sum_{j\ne i} Im\left\{\left[(\hat Y \hat Y^\dagger)_{ij}\right]^2\right\}\, f\left(\frac{|M_j|^2}{|M_i|^2}\right)
\eea
where the {\it hat} matrices are Yukawa matrices evaluated in the basis in which the Majorana mass matrix is diagonal and $M_i$ are the Majorana masses given in eq. (\ref{majei}) which, in terms of $\alpha$ and $ \phi$, can be written as:
\bea
M_1&=& |A|\,e^{i\phi_A}\,(1+\alpha \,e^{i\phi})\nn \\
M_2&=&|A|\,e^{i\phi_A} \nn \\
M_3&=& |A|\,e^{i\phi_A}\,(-1+\alpha \,e^{i\phi}).\nn
\eea 
For supersymmetric theories, the $f$-function is given by:
\bea
\label{eq:effe}
f(x)=-\sqrt{x}\left[\frac{2}{x-1}+\log \left(\frac{1+x}{x}\right)\right].
\eea
Defining $\Omega$ as the unitary matrix which diagonalizes the Majorana mass matrix,  
the Yukawa matrix in this basis is given by $\hat Y=\Omega^T\,Y_\nu$ 
and the product $\hat Y \hat Y^\dagger$ reads:
\be
 \hat Y \hat Y^\dagger = \Omega^T\, Y_\nu  Y_\nu^\dagger\, \Omega^*~.
\ee
At LO:
\bea
\Omega&=&U_{TBM}\,{\rm diag}(e^{i \phi_1},e^{i \phi_2},e^{i \phi_3})= U_{TBM}\,U_\phi \nn
\\
Y_\nu \, Y_\nu^\dagger &=& |y_\nu|^2\,I
\eea
and the $\epsilon_i$ parameters are all vanishing \cite{Jenkins:2008rb}. At the next-to-leading order, one has to take into account the corrections to the Yukawa matrix as well as to the Majorana mass matrix, which reflects in a different structure of the $\Omega$ matrix, in such a way that: 
\bea
Y_\nu&=&Y_{LO}+\delta Y \nn \\
\Omega&=&U_{TBM}\,U_\phi+\delta \Omega \nn
\eea
where both $\delta Y$ and  $\delta \Omega$ are of ${\cal O}(\varepsilon')$.
This means that the correction to the matrix product $\hat Y \hat Y^\dagger$ is given by:
\bea
\delta (\hat Y \hat Y^\dagger) =   (\delta \Omega)^T\, Y_{LO}  Y_{LO} ^\dagger\, \Omega^* +
\Omega^T\, Y_{LO}  Y_{LO}^\dagger\, (\delta \Omega)^* +  \Omega^T\, \delta(Y_\nu  Y_\nu^\dagger)\, \Omega^* ~.\nn
\eea
The first two terms do not contribute to the  $\epsilon_i$ parameters, because of the unitarity of $\Omega$, and the relevant contribution arises from the last one.
In the basis in which the charged leptons are diagonal, the NLO Yukawa matrix $\delta Y$ is determined by the following relevant operators \footnote{We omit the operator of the form $\frac{1}{\Lambda} (\nu^c \ell)_1\,\xi \,h_u$ because it develops a VEV in the same direction as the leading order Dirac operator.}
\bea
\delta W_\nu^{\rm dirac} &=& \frac{y'_\nu}{\Lambda} \left[(\nu^c \ell)_{3A}\,\varphi_S\right]\,h_u 
+\frac{y''_\nu}{\Lambda} \left[(\nu^c \ell)_{3S}\,\varphi_S\right]\,h_u ~.
\eea
Then one easily obtains:
\bea
\hat Y \hat Y^\dagger = 
\left(
\begin{array}{ccc}
 |y_\nu|^2 + 6 \, a \,\varepsilon'  &  0   & 2 \sqrt{3}\, b\, \varepsilon'\,e^{i(\phi_1-\phi_3)}\\
0 & |y_\nu|^2  & 0 \\
2 \sqrt{3}\, b\, \varepsilon'\,e^{i(-\phi_1+\phi_3)}  &   0   &  |y_\nu|^2-6 \, a \,\varepsilon'
\end{array}
\right) 
\eea
where we defined $a=\Re(y_\nu\,y^{'*}_\nu)$ and $b=\Re(y_\nu\,y^{''*}_\nu)$.
At leading order in $\varepsilon'$, the $\epsilon$ parameters are then given by:
\bea
\epsilon_1 &=&\varepsilon^{'2}\; \frac{3 \, b^2}{2\pi\, |y_\nu|^2}\,\sin\left[2(\phi_1-\phi_3)\right]\, 
f\left(\frac{|M_3|^2}{|M_1|^2}\right) \nn \\
\epsilon_2 &=&0 \\
\epsilon_3 &=&\varepsilon^{'2}\; \frac{3 \, b^2}{2\pi\, |y_\nu|^2}\,\sin\left[2(\phi_3-\phi_1)\right]\, 
f\left(\frac{|M_1|^2}{|M_3|^2}\right). \nn
\eea
A relevant feature here is that, at the perturbative order we are working, $\epsilon_2$ is vanishing.
The parameter governing leptogenesis depends on the heavy neutrino mass spectrum.
In the case of the normal hierarchy for the light neutrinos, the lightest Majorana mass is $M_3$ and leptogenesis is governed by $\epsilon_3$. Since 
\bea
\nn
f\left(\frac{|M_1|^2}{|M_3|^2}\right)\sim -3\,\frac{|M_3|}{|M_1|} \nn \\
\eea
we get:
\bea
\epsilon_3 &=&\varepsilon^{'2}\; \frac{9 \, b^2}{2\,\pi\, |y_\nu|^2}\,\sin\left[2(\phi_1-\phi_3)\right]\, 
\frac{|M_3|}{|M_1|}~.
\eea
In the normal hierarchy case $\sin\left[2(\phi_1-\phi_3)\right]\, 
\frac{|M_3|}{|M_1|}$ is typically of ${\cal O}(10^{-1})$. Thus, assuming  $3 \, b^2/\pi\, |y_\nu|^2\sim {\cal O}(1)$, with $\varepsilon \sim {\cal O}(10^{-2})$ we obtain  $\epsilon_3\sim {\cal O}(10^{-5})$, which is compatible with the requirement $\epsilon_3\sim {\cal O}(10^{-5}-10^{-6})$ needed, within large ambiguities, to reproduce the observed asymmetry.

In the  inverted hierarchy case  the largest contribution to leptogenesis comes from $\epsilon_1$.
The $f$-function assumes now a more complicated structure. By neglecting terms proportional to $r$, we get
\bea
f\left(\frac{|M_3|^2}{|M_{1}|^2}\right)\sim -\sqrt{1+2\alpha^2}\,\left[\frac{1}{\alpha^2}+\log\left(1+\frac{1}{1+2\alpha^2}\right)\right]  
\eea
and 
\bea
\sin\left[2(\phi_1-\phi_3)\right]~f\left(\frac{|M_3|^2}{|M_{1}|^2}\right)\sim \frac{2 (\alpha^2-1) \sqrt{4-\alpha^2}}{\sqrt{1+2\alpha^2}}\left[\frac{1}{\alpha}+\alpha \log\left(1+\frac{1}{1+2\alpha^2}\right)\right]~.
\eea
This function becomes very large at small $\alpha$ values but is of order 1 for $\alpha \gtrsim 1$. Still assuming $ 3 \, b^2/\pi\, |y_\nu|^2\sim {\cal O}(1)$, the requirement $\epsilon_1\sim {\cal O}(10^{-5}-10^{-6})$ can be easily fulfilled, at least in some intervals of $\alpha$ near the points $\alpha=1$ and $\alpha=2$ where the above function vanishes. 

In conclusion the present model is compatible with the constraints derived from leptogenesis.

\section{Conclusion}

We have presented and discussed an $A_4$ model for TB mixing of the see-saw type which, in spite of being based on a most economical flavour symmetry and field content, still it is phenomenologically viable. In particular TB mixing is exact in LO while all mixing angles receive corrections at higher orders. The charged lepton mass hierarchy is determined by the $A_4\times Z_4$ flavour symmetry itself without invoking a Froggatt-Nielsen  $U(1)$ symmetry, as in refs.\cite{lin1, lin2}. A value of $\theta_{13} \sim \mathcal{O}(\lambda_C^2)$ is indicated which is within the sensitivity of the experiments which are now in preparation and will take data in the near future.  This example shows once more that the results derived from $A_4$ are robust and, in particular, do not depend on the detailed mechanism that produces the hierarchy of charged lepton masses. 
In typical $A_4$ models of the see-saw type the light neutrino spectrum has the same features. Among the 3 complex masses the sum rule in eq. (\ref{sumr}) holds. This sum rule implies that the lightest neutrino has a non vanishing mass. The model is compatible with either a normal hierarchy or an inverse hierarchy spectrum. We have studied the spectrum in detail in these different cases and discussed the predictions for  the mass eigenvalues, as well as for $m_{ee}$ and leptogenesis.

\section*{Acknowledgements}

This work has been partly supported by the Italian Ministero dell'Universit\`a e della Ricerca Scientifica, under the COFIN program (PRIN 2006) and by the European Commission
under contracts MRTN-CT-2006-035505 and MRTN-CT-2004-503369.
We thank Yin Lin as well as Ferruccio Feruglio, Claudia Hagedorn and Luca Merlo for some interesting comments and discussions.


\vfill
\newpage


\begin{thebibliography}{99}

\bibitem{data1}
B.~T.~Cleveland {\it et al.},
Astrophys.\ J.\  {\bf 496} (1998) 505;
J.~N.~Abdurashitov {\it et al.}  [SAGE Collaboration],
Phys.\ Rev.\  C {\bf 60} (1999) 055801
[arXiv:astro-ph/9907113];
W.~Hampel {\it et al.}  [GALLEX Collaboration],
Phys.\ Lett.\  B {\bf 447} (1999) 127;
S.~Fukuda {\it et al.}  [Super-Kamiokande Collaboration],
Phys.\ Rev.\ Lett.\  {\bf 86} (2001) 5651
[arXiv:hep-ex/0103032];
J.~P.~Cravens {\it et al.}  [Super-Kamiokande Collaboration],
Phys.\ Rev.\  D {\bf 78} (2008) 032002
[arXiv:0803.4312 [hep-ex]];
Q.~R.~Ahmad {\it et al.}  [SNO Collaboration],
Phys.\ Rev.\ Lett.\  {\bf 87} (2001) 071301
[arXiv:nucl-ex/0106015];
S.~N.~Ahmed {\it et al.}  [SNO Collaboration],
Phys.\ Rev.\ Lett.\  {\bf 92} (2004) 181301
[arXiv:nucl-ex/0309004];
B.~Aharmim {\it et al.}  [SNO Collaboration],
Phys.\ Rev.\ Lett.\  {\bf 101} (2008) 111301
[arXiv:0806.0989 [nucl-ex]];
Y.~Fukuda {\it et al.}  [Super-Kamiokande Collaboration],
Phys.\ Rev.\ Lett.\  {\bf 81} (1998) 1562
[arXiv:hep-ex/9807003];
M.~Ambrosio {\it et al.}  [MACRO Collaboration],
Phys.\ Lett.\  B {\bf 517} (2001) 59
[arXiv:hep-ex/0106049];
M.~Apollonio {\it et al.}  [CHOOZ Collaboration],
Phys.\ Lett.\  B {\bf 466} (1999) 415
[arXiv:hep-ex/9907037];
M.~Apollonio {\it et al.}  [CHOOZ Collaboration],
Eur.\ Phys.\ J.\  C {\bf 27} (2003) 331
[arXiv:hep-ex/0301017];
F.~Boehm {\it et al.},
Phys.\ Rev.\  D {\bf 64} (2001) 112001
[arXiv:hep-ex/0107009];
K.~Eguchi {\it et al.}  [KamLAND Collaboration],
Phys.\ Rev.\ Lett.\  {\bf 90} (2003) 021802
[arXiv:hep-ex/0212021];
M.~H.~Ahn {\it et al.}  [K2K Collaboration]
Phys.\ Rev.\ Lett.\  {\bf 90} (2003) 041801
[arXiv:hep-ex/0212007];
E.~Aliu {\it et al.}  [K2K Collaboration],
Phys.\ Rev.\ Lett.\  {\bf 94} (2005) 081802
[arXiv:hep-ex/0411038];
D.~G.~Michael {\it et al.}  [MINOS Collaboration],
Phys.\ Rev.\ Lett.\  {\bf 97} (2006) 191801
[arXiv:hep-ex/0607088];
P.~Adamson {\it et al.}  [MINOS Collaboration],
Phys.\ Rev.\ Lett.\  {\bf 101} (2008) 131802
[arXiv:0806.2237 [hep-ex]].


\bibitem{data}
A.~Strumia and F.~Vissani,
arXiv:hep-ph/0606054;
G.~L.~Fogli {\it et al.},
Nucl.\ Phys.\ Proc.\ Suppl.\  {\bf 168} (2007) 341;
M.~C.~Gonzalez-Garcia and M.~Maltoni,
Phys.\ Rept.\  {\bf 460} (2008) 1
[arXiv:0704.1800 [hep-ph]];
T.~Schwetz,
AIP Conf.\ Proc.\  {\bf 981} (2008) 8
[arXiv:0710.5027 [hep-ph]];
M.~C.~Gonzalez-Garcia and M.~Maltoni,
Phys.\ Lett.\  B {\bf 663} (2008) 405
[arXiv:0802.3699 [hep-ph]];
A.~Bandyopadhyay, S.~Choubey, S.~Goswami, S.~T.~Petcov and D.~P.~Roy,
arXiv:0804.4857 [hep-ph].


\bibitem{FogliIndication}
G.~L.~Fogli, E.~Lisi, A.~Marrone, A.~Palazzo and A.~M.~Rotunno,
Phys.\ Rev.\ Lett.\  {\bf 101} (2008) 141801
[arXiv:0806.2649 [hep-ph]];
G.~L.~Fogli, E.~Lisi, A.~Marrone, A.~Palazzo and A.~M.~Rotunno,
arXiv:0809.2936 [hep-ph].


\bibitem{MaltoniIndication}
T.~Schwetz, M.~Tortola and J.~W.~F.~Valle,
New J.\ Phys.\  {\bf 10} (2008) 113011
[arXiv:0808.2016 [hep-ph]];
M.~Maltoni and T.~Schwetz,
arXiv:0812.3161 [hep-ph].


\bibitem{review}
G.~Altarelli and F.~Feruglio,
New J.\ Phys.\  {\bf 6} (2004) 106
[arXiv:hep-ph/0405048].

\bibitem{hps}
P.~F.~Harrison, D.~H.~Perkins and W.~G.~Scott,
Phys.\ Lett.\ B {\bf 530} (2002) 167
[arXiv:hep-ph/0202074];
P.~F.~Harrison and W.~G.~Scott,
Phys.\ Lett.\ B {\bf 535} (2002) 163
[arXiv:hep-ph/0203209];
Z.~z.~Xing,
Phys.\ Lett.\ B {\bf 533} (2002) 85
[arXiv:hep-ph/0204049];
P.~F.~Harrison and W.~G.~Scott,
Phys.\ Lett.\ B {\bf 547} (2002) 219
[arXiv:hep-ph/0210197];
P.~F.~Harrison and W.~G.~Scott,
Phys.\ Lett.\ B {\bf 557} (2003) 76
[arXiv:hep-ph/0302025];
P.~F.~Harrison and W.~G.~Scott,
arXiv:hep-ph/0402006;
P.~F.~Harrison and W.~G.~Scott,
Phys.\ Lett.\  B {\bf 594}, 324 (2004)
[arXiv:hep-ph/0403278].

\bibitem{TBA4}
E.~Ma and G.~Rajasekaran,
Phys.\ Rev.\ D {\bf 64} (2001) 113012
[arXiv:hep-ph/0106291];
E.~Ma,
Mod.\ Phys.\ Lett.\ A {\bf 17} (2002) 627
[arXiv:hep-ph/0203238];
K.~S.~Babu, E.~Ma and J.~W.~F.~Valle,
  Phys.\ Lett.\ B {\bf 552} (2003) 207
  [arXiv:hep-ph/0206292];
M.~Hirsch, J.~C.~Romao, S.~Skadhauge, J.~W.~F.~Valle and A.~Villanova del Moral,
  arXiv:hep-ph/0312244;
  Phys.\ Rev.\  D {\bf 69} (2004) 093006
  [arXiv:hep-ph/0312265];
E.~Ma,
  Phys.\ Rev.\ D {\bf 70} (2004) 031901;
  Phys.\ Rev.\ D {\bf 70} (2004) 031901
  [arXiv:hep-ph/0404199];
  New J.\ Phys.\  {\bf 6} (2004) 104
  [arXiv:hep-ph/0405152];
  arXiv:hep-ph/0409075;
S.~L.~Chen, M.~Frigerio and E.~Ma,
  Nucl.\ Phys.\  B {\bf 724} (2005) 423
  [arXiv:hep-ph/0504181];
E.~Ma,
  Phys.\ Rev.\  D {\bf 72} (2005) 037301
  [arXiv:hep-ph/0505209];
M.~Hirsch, A.~Villanova del Moral, J.~W.~F.~Valle and E.~Ma,
   Phys.\ Rev.\  D {\bf 72} (2005) 091301
   [Erratum-ibid.\  D {\bf 72} (2005) 119904]
   [arXiv:hep-ph/0507148].
K.~S.~Babu and X.~G.~He,
  arXiv:hep-ph/0507217;
E.~Ma,
  Mod.\ Phys.\ Lett.\ A {\bf 20} (2005) 2601
  [arXiv:hep-ph/0508099];
A.~Zee,
  Phys.\ Lett.\ B {\bf 630} (2005) 58
  [arXiv:hep-ph/0508278];
E.~Ma,
  Phys.\ Rev.\  D {\bf 73} (2006) 057304
  [arXiv:hep-ph/0511133];
X.~G.~He, Y.~Y.~Keum and R.~R.~Volkas,
  JHEP {\bf 0604} (2006) 039
  [arXiv:hep-ph/0601001];
B.~Adhikary, B.~Brahmachari, A.~Ghosal, E.~Ma and M.~K.~Parida,
  Phys.\ Lett.\ B {\bf 638} (2006) 345
  [arXiv:hep-ph/0603059];
E.~Ma,
  Mod.\ Phys.\ Lett.\  A {\bf 21} (2006) 2931
  [arXiv:hep-ph/0607190];
  Mod.\ Phys.\ Lett.\  A {\bf 22} (2007) 101
  [arXiv:hep-ph/0610342];
L.~Lavoura and H.~Kuhbock,
  Mod.\ Phys.\ Lett.\  A {\bf 22} (2007) 181
  [arXiv:hep-ph/0610050];
S.~F.~King and M.~Malinsky,
  Phys.\ Lett.\  B {\bf 645} (2007) 351
  [arXiv:hep-ph/0610250];
S.~Morisi, M.~Picariello and E.~Torrente-Lujan,
  Phys.\ Rev.\  D {\bf 75} (2007) 075015
  [arXiv:hep-ph/0702034];
M.~Hirsch, A.~S.~Joshipura, S.~Kaneko and J.~W.~F.~Valle,
   Phys.\ Rev.\ Lett.\  {\bf 99}, 151802 (2007)
   [arXiv:hep-ph/0703046].
F.~Yin,
  Phys.\ Rev.\  D {\bf 75} (2007) 073010
  [arXiv:0704.3827 [hep-ph]];
F.~Bazzocchi, S.~Kaneko and S.~Morisi,
  JHEP {\bf 0803} (2008) 063
  [arXiv:0707.3032 [hep-ph]].
F.~Bazzocchi, S.~Morisi and M.~Picariello,
  Phys.\ Lett.\  B {\bf 659} (2008) 628
  [arXiv:0710.2928 [hep-ph]];
M.~Honda and M.~Tanimoto,
  Prog.\ Theor.\ Phys.\  {\bf 119} (2008) 583
  [arXiv:0801.0181 [hep-ph]];
B.~Brahmachari, S.~Choubey and M.~Mitra,
  Phys.\ Rev.\  D {\bf 77} (2008) 073008
  [Erratum-ibid.\  D {\bf 77} (2008) 119901]
  [arXiv:0801.3554 [hep-ph]];
F.~Bazzocchi, S.~Morisi, M.~Picariello and E.~Torrente-Lujan,
  J.\ Phys.\ G {\bf 36} (2009) 015002
  [arXiv:0802.1693 [hep-ph]];
B.~Adhikary and A.~Ghosal,
  Phys.\ Rev.\  D {\bf 78} (2008) 073007
  [arXiv:0803.3582 [hep-ph]];
M.~Hirsch, S.~Morisi and J.~W.~F.~Valle,
  Phys.\ Rev.\  D {\bf 78} (2008) 093007
  [arXiv:0804.1521 [hep-ph]].
Y.~Lin,
  arXiv:0804.2867 [hep-ph].
P.~H.~Frampton and S.~Matsuzaki,
  arXiv:0806.4592 [hep-ph];
  C. ~Csaki, C.~ Delaunay, C. ~Grojean, Y.~Grossman
   arXiv:0806.0356 [hep-ph];
F.~Feruglio, C.~Hagedorn, Y.~Lin and L.~Merlo,
  arXiv:0807.3160 [hep-ph];
F.~Bazzocchi, M.~Frigerio and S.~Morisi,
  arXiv:0809.3573 [hep-ph];
W.~Grimus and L.~Lavoura,
  arXiv:0811.4766 [hep-ph];
S.~Morisi,
  arXiv:0901.1080 [hep-ph];
P.~Ciafaloni, M.~Picariello, E.~Torrente-Lujan and A.~Urbano,
  arXiv:0901.2236 [hep-ph];
   M.~C.~Chen and S.~F.~King,
  arXiv:0903.0125 [hep-ph].

\bibitem{AFextra}
G.~Altarelli and F.~Feruglio,
  Nucl.\ Phys.\ B {\bf 720} (2005) 64
  [arXiv:hep-ph/0504165].


\bibitem{AFmodular}
G.~Altarelli and F.~Feruglio,
  Nucl.\ Phys.\  B {\bf 741} (2006) 215
  [arXiv:hep-ph/0512103].


\bibitem{AFL}
G.~Altarelli, F.~Feruglio and Y.~Lin,
  Nucl.\ Phys.\  B {\bf 775} (2007) 31
  [arXiv:hep-ph/0610165].
  
\bibitem{AFH}
  G.~Altarelli, F.~Feruglio and C.~Hagedorn,
  JHEP {\bf 0803} (2008) 052
  [arXiv:0802.0090 [hep-ph]].


\bibitem{continuous}
S.~F.~King,
  JHEP {\bf 0508} (2005) 105
  [arXiv:hep-ph/0506297];
I.~de Medeiros Varzielas and G.~G.~Ross,
  arXiv:hep-ph/0507176.
I.~de Medeiros Varzielas, S.~F.~King and G.~G.~Ross,
  Phys.\ Lett.\  B {\bf 644} (2007) 153
  [arXiv:hep-ph/0512313].
I.~de Medeiros Varzielas, S.~F.~King and G.~G.~Ross,
  Phys.\ Lett.\  B {\bf 648} (2007) 201
  [arXiv:hep-ph/0607045].
S.~F.~King and M.~Malinsky,
  JHEP {\bf 0611} (2006) 071
  [arXiv:hep-ph/0608021].
S.~Antusch, S.~F.~King and M.~Malinsky,
  JHEP {\bf 0806} (2008) 068
  [arXiv:0708.1282 [hep-ph]].


\bibitem{others}
For others approaches to the tri-bimaximal mixing see:
J.~Matias and C.~P.~Burgess,
  JHEP {\bf 0509} (2005) 052
  [arXiv:hep-ph/0508156];
S.~Luo and Z.~z.~Xing,
  arXiv:hep-ph/0509065.
W.~Grimus and L.~Lavoura,
  arXiv:hep-ph/0509239.
F.~Caravaglios and S.~Morisi,
  arXiv:hep-ph/0510321.
I.~de Medeiros Varzielas, S.~F.~King and G.~G.~Ross,
  Phys.\ Lett.\  B {\bf 648} (2007) 201
  [arXiv:hep-ph/0607045].
F.~Feruglio, C.~Hagedorn, Y.~Lin and L.~Merlo,
  Nucl.\ Phys.\  B {\bf 775} (2007) 120
  [arXiv:hep-ph/0702194].
M.~C.~Chen and K.~T.~Mahanthappa,
  Phys.\ Lett.\  B {\bf 652} (2007) 34
  [arXiv:0705.0714 [hep-ph]];
P.~H.~Frampton and T.~W.~Kephart,
  JHEP {\bf 0709} (2007) 110
  [arXiv:0706.1186 [hep-ph]];
G.~J.~Ding,
  arXiv:0803.2278 [hep-ph];
F.~Bazzocchi and S.~Morisi,
  arXiv:0811.0345 [hep-ph];
P.~H.~Frampton and S.~Matsuzaki,
  arXiv:0902.1140 [hep-ph].

\bibitem{bmm}
F.~Bazzocchi, L.~Merlo and S.~Morisi,
  arXiv:0901.2086 [hep-ph].
F.~Bazzocchi, L.~Merlo and S.~Morisi,
  arXiv:0902.2849 [hep-ph].

\bibitem{lin1}
Y.~Lin,
Nucl.\ Phys.\  B {\bf 813}, 91 (2009)
[arXiv:0804.2867 [hep-ph]].

\bibitem{lin2}
L.~Yin,
arXiv:0903.0831 [hep-ph].

\bibitem{GALT} G. Altarelli, Proceedings of *Venice 2007, Neutrino telescopes* pages 139-161  
[arXiv:0705.0860 [hep-ph]]; G. Altarelli, [arXiv:0711.0161 [hep-ph]].

\bibitem{Chankowski:2001mx}
P.~H.~Chankowski and S.~Pokorski,
[arXiv:hep-ph/0110249].

\bibitem{Kuo:2001ha}
T.~K.~Kuo, J.~T.~Pantaleone and G.~H.~Wu,
Phys.\ Lett.\  B {\bf 518}, 101 (2001)
[arXiv:hep-ph/0104131].

\bibitem{Jenkins:2008rb}
E.~E.~Jenkins and A.~V.~Manohar,
Phys.\ Lett.\  B {\bf 668}, 210 (2008)
[arXiv:0807.4176 [hep-ph]].


\end{thebibliography}
\end{document}